**Trans-NIH/Interagency Workshop on the Use and Development of Assistive Technology for the Aging Population and People with Chronic Disabilities**

**Computing Community Consortium**
**February 17, 2015**[1]

**Elizabeth Mynatt**
Professor of Interactive Computing, Georgia Tech

**Alice Borrelli**
Director of Global Healthcare Policy, Intel

**Sara Czaja**
Professor, Psychiatry and Behavioral Sciences & Industrial Engineering, University of Miami

**Erin Iturriaga**
National Institutes of Health, National Heart, Lung and Blood Institute

**Jeff Kaye**
Professor of Neurology and Biomedical Engineering, OHSU

**Wendy Nilsen**
National Institutes of Health, Office of the Director

**Dan Siewiorek**
Buhl University Professor of ECE and Computer Science, CMU

**John Stankovic**
Professor, Computer Science, University of Virginia

**Executive Summary**
As baby boomers age, the nation's elderly population continues to grow. A majority of these individuals will continue to live in their own home. Meeting this societal need requires a new generation of research that addresses the complexity of supporting the quality of life and independence of a vast, diverse and aging population. New technologies could potentially allow older adults and people with disabilities to remain in their homes longer, reduce health care costs, enhance their quality of life, and provide needed support to independent caregivers.

In September 2014, the Computing Community Consortium and National Institutes of Health National Heart, Lung, and Blood Institute held a workshop to explore the use and development of these technologies. Here we describe the outcomes of the workshop, highlighting the critical research needed to meet the needs of our aging population and providing specific recommendations for these research investments.

---

[1] Contact: Ann W. Drobnis, Director, Computing Community Consortium (202 266-2936; adrobnis@cra.org).

## 1.0 Aging in Place: A National Imperative

The first baby boomer born in 1946 turned 65 in 2011 and the last baby boomer will turn 65 in 2029. By then, the total U.S. population over 65 is projected to be 71.5 million[2] (compared with 44.6 million in 2013)[3]. The current (2014) median cost of a nursing home is $226 a day ($82,490 per year), while assisted living is $3500 a month ($42,000 per year)[4]. While the elderly population continues to get larger and costs will continue to rise, nearly ninety percent (90%) of people want to grow old in their own home and community and remain out of the hospital, nursing home or other institutional setting[5].

New technologies could potentially allow older adults and people with disabilities to remain in their homes longer, reduce health care costs and enhance the quality of life. As a response to the 2012 Institutes of Medicine Workshop, *Fostering Independence, Participation, and Healthy Aging Through Technology*[6], a multi-agency group led by the National Institutes of Health's (NIH) Heart, Lung, and Blood Institute (NHLBI) was created to investigate innovative in-home monitoring technologies to enhance health while aging in place (e.g., BP monitoring, medication adherence aids). In a healthcare environment where decision-making is increasingly moving toward evidence-based, there is a much greater need for, published, quantitative information on the effectiveness of in home technologies for use by providers and patients.

While much has been written about the concerns and anticipated costs of caring for an aging population, there is a glaring need for sufficient guidance for the next generation of research aimed at supporting the quality of life and independence of aging adults. Moreover, care for this population includes addressing the needs of adults living with chronic disease and disabilities.

We are not starting with a blank slate. More than two decades of research points to the feasibility of making complex technologies useable for older adults and people caring for them, both professional and family; creating monitoring systems that recognize components of everyday activities and age-related declines in performance; and creating systems that augment and amplify human abilities, ranging from cognitive and physical supports to systems that enhance social connectedness and meaningful participation in community life.

However there is a growing realization that a new generation of research is needed to translate these gains into meaningful social and economic contributions for aging adults, and that this research must grapple with complex challenges inherent in the people, conditions, and technologies for successful aging. The spectrum of older adults who could benefit from healthcare technologies in the home ranges from working adults in their 50s to a growing population of centenarians. While it is obvious that the capabilities and needs of the aging population vary tremendously, it is also important to recognize that the capabilities of an individual can vary significantly from day to day, and that these capabilities will decline (or not) at different rates.

---

[2] http://www.census.gov/prod/2014pubs/p25-1141.pdf
[3] http://quickfacts.census.gov/qfd/states/00000.html
[4] https://www.genworth.com/dam/Americas/US/PDFs/Consumer/corporate/131168-032514-Executive-Summary-nonsecure.pdf
[5] www.aarp.org
[6] http://www.iom.edu/Reports/2013/Fostering-Independence-Participation-and-Healthy-Aging-Through-Technology.aspx

Many health conditions are correlated with aging. Effective home management of such chronic diseases as dementia, heart failure, hypertension, chronic pulmonary disease, and arthritis would reduce hospitalizations and other healthcare costs while also improving quality of life. Again complexity abounds. Older adults are more likely to be diagnosed with multiple chronic diseases, take multiple medications, and are more likely to have physical and cognitive impairments.

Systems of care, both traditional healthcare as well as informal wellness and family caregiving, come with their own requirements for successful support. Traditional healthcare is dominated by a system of specified protocols of care, specialized equipment, and reimbursement models for costs. This system is seemingly at odds with general strategies for wellness, consumer technology and yet-to-be created economic and business models. Informal and family caregiving raises additional issues for caregiver training and burnout as well as economic costs for lost productivity. All systems of care raise privacy considerations.

Given the economic concerns about the costs of healthcare and disease management, it is easy to forget that the priorities of older adults and their families revolve around health, wellness, independence and quality of life. This point is not simply rhetorical, but points to the need for a research agenda that empowers meaningful and purposeful life. Systems that reflect these priorities have a greater likelihood of adoption and long-term impact.

In summary, there is a need for a new generation of research that addresses the complexity of supporting the quality of life and independence of a vast, diverse, and aging population. While there are common themes and needs in this research that we describe shortly, we must start by recognizing that there is more than one needed path and approach to meet these diverse needs. One path includes the tight integration of chronic disease management in the home with existing acute healthcare systems. Another path embraces comprehensive home health for improving nutrition and social connectedness while combating physical, cognitive and psychological ailments. Yet another path emphasizes wellness, consumer technologies and removing basic barriers to meaningful community participation. These paths will intersect in interesting ways for individuals, families, healthcare providers, and communities. However research is critically needed to illuminate these paths and to make measurable strides in our care and support for over 15% of our nation's citizens.

## 2.0     Multi-disciplinary, Multi-Agency Collaboration Required

Currently, there is a paucity of research in aging in place technologies (AiPT) from a systems approach that includes the expertise of health and computer science researchers, the expertise and capabilities of the many agencies that address the needs of an aging population, and the policy landscape that governs medical and technology interventions. This paper summarizes the recommendations of a Trans-NIH/Interagency Workshop on the "Use and Development of Assistive Technology for the Aging Population and People with Chronic Disabilities," convened by The National Institutes of Health National Heart, Lung, and Blood Institute and the Computer Research Association's Computing Community Consortium (CCC).

The overarching goal of the workshop was to bring together needed interdisciplinary expertise, assess the state of the science at the human, medical, and technology levels, and articulate a research vision for a systems engineering approach to the development of technologies and solutions to support the home management of persons with significant chronic diseases and their family care providers.

Over 70 people participated in this two-day workshop. Participating agencies included the National Institutes of Health (NIH), National Science Foundation (NSF), Centers for Medicare & Medicaid Services (CMS), U.S. Department of Housing and Urban Development (HUD), U.S. Department of Health & Human Services (HHS), U.S Food and Drug Administration (FDA), U.S. Department of Veterans Affairs (VA), U.S. Department of Education (Ed), The National Academies (NAS), The Federal Communications Commission (FCC), and National Institute on Disability and Rehabilitation Research (NIDRR). Researchers from across the nation brought together expertise from academic and industry projects including Oregon Center for Aging & Technology (ORCATECH)[7], Senior Independent Living Research (SILvR) Network[8], Tiger Place[9], Center for Research and Education on Aging and Technology Enhancement (CREATE)[10], Quality of Life Technology Center (QoLT)[11], Georgia Tech's Aware Home[12], VA Geriatrics and Extended Care[13], ElderTree[14] and the Henry Ford Health System[15].

This workshop report is a concrete step to providing a research agenda needed in the development and application of technology to home management of chronic diseases. Major areas addressed by the workshop included:

- Critical success factors to help the aging or disabled stay in their homes especially in low resource and underserved populations.
- Gaps in research for technologies and systems that enable home care.
- Mechanisms by which health researchers, ethnographers, computer scientists and multidisciplinary research teams can effectively partner and address the next generation of systems to support home care for elders.
- Recommendations for addressing those critical gaps identified throughout the workshop discussions.
- Policy barriers to patients needing access to the enabling technologies.

The scope of the discussions centered on four main challenges: designing for the population, sensing innovations required to enhance health, using technology to identify and support transitions in health, and utilizing the new non-health technologies to support health in smart homes.

### 3.0     The Complex Needs of Older Adults

Technology has a great deal to offer in terms of enhancing independence and the quality of life for older adults and their families. This promise is especially true today as healthcare for those with both acute and chronic conditions is increasingly occurring in community settings such as the home rather than in professional medical settings. Thus a wide array of individuals including people who may care for themselves, and those providing care, who may be professional or family caregivers, are performing a complex array of tasks and engaging in procedures that were **previously performed by healthcare professionals** and using a variety of equipment and technologies. A critical issue is to ensure that the technologies are safe, useful and usable to these diverse user groups. In addition, it is equally important to achieve a balance between technology support, augmentation and social connectedness and to strive to maintain human dignity, privacy and safety. Further, technology solutions must maximize the potential benefits that technology has to offer without overwhelming the user.

---

[7] http://www.ohsu.edu/xd/research/centers-institutes/orcatech/index.cfm
[8] http://silvrnetwork.org/
[9] http://www.americareusa.net/location/tigerplace/
[10] http://www.create-center.org/
[11] http://www.cmu.edu/qolt/index.html
[12] http://www.awarehome.gatech.edu/
[13] http://www.va.gov/geriatrics/
[14] http://eldertreecare.com/
[15] http://www.henryford.com/

Older adults represent a highly diverse user group with varying needs, capabilities and preferences. At the outset it is important to recognize that life expectancy has increased due to changes in lifestyle behaviors and advances in medical technologies and there are increasing numbers of people who are living into their 80s and 90s and beyond. Recently this latter cohort has been referred to as the "oldest old" (85+ years). In fact, since 1980 the number of people in the United States aged 100 or above has increased by 83%. A person at 60 or 70 years of age is typically very different from someone in their 80s; there are also differences between those who are 80-89 and those 90+ years. The older adult population is also becoming increasing ethnically diverse and by 2030 28% of the 65+ population will be ethnic minorities with large growth in the Hispanic and African American groups. The living arrangements of older adults also varies considerably; 57% of older adults live with a spouse, 28% live alone and about three percent live in senior housing facilities. The likelihood of living alone is greater for women and those in the older cohorts. Importantly, a substantial number of older people also live in areas that are rural or outside of major urban areas. This often results in reduced access to needed resources and services. Older adults also vary greatly in terms of their health and functional status.

We can view human/technology interactions as components of a system which involves the user populations who have varying characteristics, capabilities and limitations; the tasks that they are engaged in which have an inherent set of demands; the equipment or technologies that they are using in the performance of these tasks which also have an inherent set of demands; and the environment, physical and social in which these interactions occur (Figure 1). A critical component of this system is the user in this case, which largely consists of the older adult and their care providers.

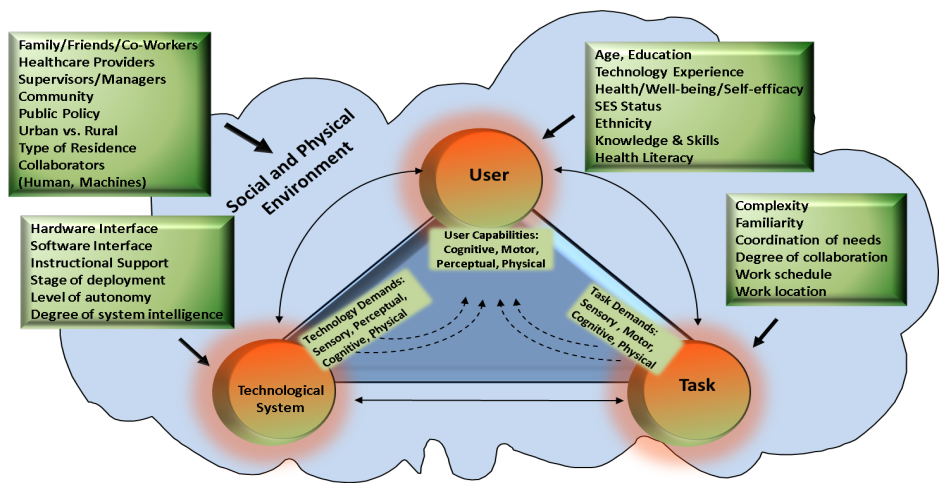

**Figure 1. CREATE Model of the Human/Technical System**

The development of technologies and solutions that meet these design requirements and support successful "aging in place" for both healthy older adults, those with significant chronic diseases and their family care providers requires a **user centered design approach**. Basic tenets of this approach including "knowing" the user population and involving them early and throughout the design process and using an iterative design approach. Thus a fundamental issue for designers is the understanding of the characteristics, needs, capabilities and preferences of older adults as well as caregiver populations that include both family caregivers and healthcare providers. This user-centered design approach extends past the goals of designing systems that compensate for age-related declines in physical strength and sometimes cognitive capacity; effective design should also focus on amplifying and reinforcing retained abilities in older adults.

As noted, older adults are part of a social and care network that includes family caregivers as well as healthcare professionals such as physicians, nurses, home health workers, care managers and varying types of therapists There are numerous ways that technology can be used to foster the interactions within these networks and facilitate care coordination. For example, home sensing and monitoring systems can be used to track changes in activity patterns of an older adults or safety events such as falls. This information can be helpful to family members who live in distant locations or work as well as to care managers. Healthcare providers can also use these types of systems to monitor the health status of both the older adult and family caregiver. Thus designers of monitoring and sensing systems need to understand the characteristics, needs and preferences of a broad array of users. In this example, from the perspective of the older adult a concern might be privacy and data sharing. From the perspective of the family caregiver, concerns may relate to amount and usability of the data, and from the healthcare providers, concerns might center around increased workload demands and protocols for communicating with family members. Further all of these interactions take place within living and work environments that vary greatly across individuals.

Open research, design and engineering challenges include:

**Adaptable systems**: Creating systems that assess and adapt dynamically to changes in a person's capabilities both in the short term (e.g. good days and bad days) and in the long term as some capabilities decline.

**Human-Centric Design**: Designing systems that meet people where they are in terms of their knowledge and priorities in addition to their cognitive and physical capabilities. For example systems could leverage older conceptual models of communication technologies and should frame engagement and activities based on overall quality of life goals.

**Care-Networks**: Designing systems that support informal and formal caregivers and care networks. These systems must dynamically manage care coordination and information sharing while balancing privacy and responsibilities between a person and his or her care network.

**Future-Proofing**: Future proofing current technology innovations so that newly invented technologies will be usable by the next generation of older adults.

**Safe and Reliable Devices:** Home medical devices must work reliably, in a wide variety of conditions and locations, and be useable by non-experts, some with physical and visual difficulties**.**

**4.0     Designing for Health Transition Trajectories**

To be most effective, technologies to facilitate aging in place must fit into the fabric of life in contrast to isolated and specialized acute healthcare. Technology design needs to acknowledge that the major determinants of aging in place are different depending on where one stands in the trajectory of change in life circumstance and health status. A useful framework for considering this dynamic aspect of independence is to consider the transition phases from being well to being at risk (by virtue of emergence of minor symptoms or risk factors) and then on to having manifest disease or disability. Each phase requires specific preventions that may be put in place to mitigate the transition to unwanted care scenarios or living situations.

These three phases in the trajectory of health translate into opportunities for prevention across a continuum: primary prevention (promoting health and well being when well), secondary prevention (providing early intervention when risk is identified) and tertiary prevention (disease management when illness becomes chronic).

Technology has an important role to play across this continuum. There are many opportunities: an early warning system of "ambient independence measures" composed of activity sensors, socialization indicators and health data for primary prevention of loss of independence [NIA AG042191]; a tele-presence robot to provide secondary prevention for isolated persons living alone [Seelye, 2012]; or delivering daily video chats to isolated seniors to increase social engagement and prevent cognitive decline [NIA AG033581].

More leading-edge pervasive computing implementations using remote monitoring to prevent persons at risk from declining or losing independence have been established. These programs include the Oregon Center for Aging & Technology's (ORCATECH) Life Laboratory cohort [Kaye, 2011], Center for Advanced Studies in Adaptive Systems (CASAS) [Crandall, 2012], and Tiger Place [Rantz, 2013], among others. These programs have shown that continuous, high frequency remotely sensed activity data can detect meaningful changes in function such as in mobility [Kaye, 2012], cognition [Kaye, 2013] medication taking ability [Hayes, 2009] or sleep behaviors [Hayes, 2014].

Here we discuss two major health transition trajectories - cognitive and physical decline – and the ways in which AiPT can play a role within the primary, secondary and tertiary prevention framework.

**Cognitive Decline:** Cognitive impairment in older adults has been associated with numerous negative health care consequences: increased health care utilization, placement in long-term care facilities, number of days in the hospital, falls, loss of self-esteem, poorer quality of life (also for caregiver), conversion to dementia, and morbidity and mortality[16].

How can we design technologies at all stages of prevention to reduce and delay the impact of cognitive aging and dementia on everyday functioning?

**Primary prevention** focuses on supporting brain health through: exercise, cognitive engagement, social engagement, good eating habits, good sleep hygiene, and stress reduction. Technologies to support brain health include: wearable systems for tracking mobility and other health related activities; social networking technologies; and intelligent systems that can learn and monitor behaviors and prompt to assist in increasing better health related behaviors.

---

[16] http://www.cdc.gov/aging/pdf/cognitive_impairment/cogImp_poilicy_final.pdf

Open challenges include how can technologies be designed and used to engage individuals in healthy behaviors, motivate continued engagement in behavior that support brain health, and introducing brain health at an early age by making technology fun and a status symbol.

**Secondary prevention** involves early detection of emergent cognitive changes through intelligent systems that monitor behaviors and detect changes that suggest deviations in a person's health – both acute and gradual.

Open challenges focus on making continuous monitoring acceptable and reliable. Monitoring systems may easily result in large amounts of data to be stored and interpreted. Algorithms must detect low base-rate events from sensor data and critical situations while avoiding a high false positive rate. Moreover these monitoring systems must prevent information overload for users and utilize reliable and sustainable sensing technologies.

**Tertiary prevention** can complement formal human care and reduce excess morbidity. Such technologies should increase a sense of safety and independence (including personal emergency response systems), increase confidence in performing everyday activities, allow adults to feel more active in their care, have a positive impact on quality of life, decrease feelings of isolation, and improve communication with loved ones and improve social support. Technologies can promote safety, foster social communication, act as a memory enhancer, and support daily activities. One such technology could be an intelligent prompter. Questions to be answered include when to prompt and what type of cues. Tertiary prevention should enhance the quality of life for caregivers by providing emotional support and information, decrease worry and burden, and decrease additional stressors. Professional providers can gain by having at the point of care objective real-time records of key information for effective cognitive or dementia care such as a person's overall activity levels, sleep, weight and medication adherence.

Open challenges include designing intelligent context-aware technologies that are passive – requiring no or minimal user initiation or maintenance – yet providing useful and usable assistance.

**Physical Decline:** There is a growing need to provide support for temporary and permanent physical declines in the aging population. By 2030, 4% of the population will experience a stroke at a cost of over $180 billion. In 2004, there were 450,000 total knee replacements and 230,000 total hip replacements[17]; in 2006, 250,000 rotator cuff surgeries[18] and in 2009, 250,000 anterior cruciate injuries[19]. Six or more months of rehabilitation are commonly required.

**Primary prevention** in this context focuses on maintaining physical function. In the older population this is particularly important for the prevention of falls. Physical home modifications, such as switching out bathtubs for barrier free showers, have proved to be effective[20]. Reducing cognitive decline (discussed above) and managing medications for example through automated medication dispensers and alerts to pharmacies can also ultimately contribute to preventing physical injury.

---

[17] American Academy of Orthopedic Surgeons (AAOS). The burden of musculoskeletal diseases in the United States. 2nd ed. Rosemont (IL): American Academy of Orthopedic Surgeons; 2008.
[18] Colvin AC, Egorova N, Harrison AK, Moskowitz A, Flatow EL. National trends in rotator cuff repair, Journal Bone Joint Surg Am., 2012; 94:227–233.
[19] Collins, S.L., Van Valin, S.E., Anterior Cruciate Ligament Tear in a 7-Year-Old Athlete, The American Journal of Orthopedics, www.amjorthopedics.com, January 2013, 33-36.

**Secondary prevention** in the context of physical function involves early detection, monitoring, and temporary assistance for example following identification of risk for falling. Such technologies as smart orthotics, canes or walkers to provide improved sensory feedback and support can play a role. On-line video feed-back exercise programs can help enhance balance and strength.

**Tertiary prevention** again complements formal human care. Increasingly sophisticated prosthetic tools as well as robotic assistants can bridge the gap between human ability and the performance of many daily tasks.

Overall designing for health transition trajectories requires a suite of approaches for weaving technical assistance into the fabric of daily life. The trajectories described shed light on the combination of approaches that will undoubtedly comprise aging in place. The consumer market may dominate primary prevention although this market will be well suited to anticipate the shifts between primary and secondary prevention. Secondary prevention may often include extensions to acute care and disease management, but again the home setting is not limited to those concerns. Tertiary prevention shifts into more comprehensive home health approaches. Given the extent of human need, solutions for tertiary prevention may often require prior experience with less extensive, secondary and primary prevention approaches, especially with respect to cognitive decline, for those solutions to be effectively adopted and utilized.

In total, this area presents numerous open research, design, and engineering challenges including:

**Engagement:** Engaging users effectively is an open challenge for many home care scenarios. Therapeutic applications often fail to motivate users for an effective period of time. One approach is to effectively personalize the user experience rendering it more relevant and more psychologically appealing. Reducing unneeded complexity and making systems easier to use across the board will help decrease abandonment and increase adoption of systems that can have long term benefits.

**User Modeling:** Effective user modeling is required to realize many monitoring and engagement goals for aging in place technology. The inter- and intra-person variability of users challenges the balance of creating economically viable systems that meet the needs of a large population while, at the same time, fine tuning each system to match the behavior, capabilities and needs of a specific individual who may also dramatically change in his/her abilities.

**Privacy and User Acceptance:** Designing for trajectories of care draws attention to the need for designing for trajectories of privacy concerns and user acceptance. Primary prevention approaches will likely need to emphasize privacy and local control of information while tertiary prevention will likely require greater tradeoffs between privacy, local control and overall independence and quality of life.

---

[20] Sanford, J.A. (2010). Physical Environment and Home Healthcare. In National Research Council Committee on the Role of Human Factors in Home Healthcare, *Role of Human Factors in Home Healthcare*, Washington, DC: National Academies Press.

**5.0 Innovation Needed: Sensing, Actuation and System Integration Technology**

Many technologies and smart monitoring systems have been deployed in homes and assisted living facilities to support aging in place. While these deployments are beginning to show the potential of technology to support aging in place, many technical research questions must still be solved. It is projected that solving these problems would produce enormous gains for the utility and lower cost of such systems. In this section, a few examples of some of the key open technical research questions are discussed.

Many aging in place monitoring systems are based on detecting activities of daily living (ADLs) and then identifying anomalies in behaviors. Upon detecting an anomaly, messages or alarms are sent to caregivers or family members. The logic in these current detection systems is typically limited in one or more important ways, including: the complexities of human behaviors and the complexities of the environment in the home or assisted living complex are underestimated. These limitations often result in inaccurate assessment of ADLs and subsequently inaccuracies of what is anomalous behavior. Research is required to build more accurate models of human behaviors that address differences in behaviors based on seasons, days of week, with or without visitors, before or after tragic events, or new medical conditions or operations, frequency of behaviors such as twice a week, or once a month, etc. Better use of correlations among behaviors and detection of causality are necessary.

Long-term deployments in open environments such as homes and assisted living facilities also cause reliability and robustness problems that must be solved. The reliability of the sensing must account for much more than a failure of a specific device. For example, sensors may be placed on furniture and assumptions made about where that furniture is located. A movement fault occurs if it is moved to another room. Movement of household plants and furniture may also block sensor signals, creating other faults. Acoustic sensors can suffer from an enormous number of ambient sounds. All of these environmental conditions give rise to inaccurate assessments of ADLs. Research is needed for more robust solutions that account for long term, evolving, and open environments. Too many current solutions work only under a very constrained and rigid set of assumptions about who, when, where and why a device is used.

While aging in place systems are currently being deployed, there is still a significant research need to remove many burdensome aspects of some current systems. For example, some of these systems are difficult and time consuming to deploy, costly, and may be difficult to use, especially for older adults who might be frail, have significant chronic conditions, or suffer from poor hearing and eyesight. The physical appearance and obtrusiveness of devices may also limit acceptance of these systems. Once deployed, another critical issue arises in that these systems also may be difficult to maintain. Power, networking, sensor faults, and non-expert users contribute to maintenance difficulties. New research in fault detection, automatic and remote repair, reliable communications, and energy scavenging would improve the situation. In general, having the sensor systems be minimally obtrusive, ideally disappearing into the background is an important research goal.

Overall, aging in place systems must be trustworthy and maintain privacy. New research is needed to enable these systems to operate in the presence of malware, spyware, jamming, and other security and privacy attacks both within the home and in the cloud where much of the data resides. The open environments where these systems operate also give rise to the possibility of physical attacks on the devices and systems. Trustworthiness includes, not only the devices and controlling software, but also the data analytics, decision making, and the incorporation of accurate medical knowledge and expertise.

It is projected that there will be an extremely diverse set of people using aging in place systems. Diversity arises from different education and SES levels, ethnicity, (multiple) chronic medical conditions, poor vision and hearing, location, and physical frailty. Consequently, significant research is required on the human-computer interaction (HCI) aspects of aging in place devices and systems. These interfaces must be developed with the goal of making data actionable. The entire path from data collection, data integration, and knowledge creation to decision making must occur and be facilitated by the human-computer interfaces. Interfaces must also be trustable and useable even as the systems and problems grow in complexity. New research is required to make interfaces intention aware so that users of all types are led to the proper information, aids, and interventions. Personalization and dynamic adaptability are also required for the interfaces as baseline behaviors of individuals change over time. In the end, interfaces must be effective at helping people make informed decisions regardless of education and poverty levels, ethnicity, chronic conditions, etc. Security and privacy issues complicate the search for effective HCI solutions and are additional open research questions.

In total, this area presents numerous open research, design, and engineering challenges including:

**Monitoring:** While monitoring is an assumed piece of many "aging in place" scenarios, tremendous progress still needs to be made in this space. Monitoring systems must assume noisy data stemming from the complexity of the home environment and the limitations of most sensors. Almost all scenarios require very low false positive rates and fusing data from heterogeneous sources. Monitoring for clinical applications must contend with a lower data quality compared to more controlled, and likely more expensive, acute care settings.

**System Resilience:** Many household appliances operate for a decade or more in contrast to the churn of consumer electronics. Aging in place technologies will need to provide utility over long periods of time, with minimal configuration, in a dynamic environment. To do so systems will need to rely on long-lasting, or power harvesting batteries. They will need to accommodate the addition and subtraction of heterogeneous components. They will also need to accommodate fluctuations in the environment, from shifts in ambient lighting and noise to rearranged furniture.

**Predictive and Decision Analytics**: Information produced by aging in place technologies will need to be comprehensible and actionable for a large set of users; ranging from older adults and other home occupants, informal and formal caregivers, clinical providers and payers. Predictive analytics will need to depict actionable trends while decision analytics will need to prioritize possible actions in a care network.

### 6.0 Barriers to Actionable Progress

We identified a set of barriers that must be addressed in order to make actionable progress to meeting the needs of our aging population through innovations in home health technologies:

**The need to better understand the target users**

It is important to understand the needs, preference, and context (home environment) of the target users in order to design and adapt technologies that take advantage of the strengths of the target populations and meet the necessary requirements to facilitate healthy independence. Once the target user is understood, technologies can be developed to motivate user engagement in social interactions, healthy behaviors, and allow them to becoming more aware and responsible for their own health.

**The need for actionable evidence**

In the rapidly changing world of technology, it is important that all technology provide timely, personalized, actionable information, with reliable interfaces and systems to support evidence-based decision-making and follow best practices for design and implementation. There are many ways to create evidence. Most important is to design with a highly focused question and the input of all end users whether they are seniors, clinicians or technologists. Evidence may come from a variety of sources: expert panel, case series, randomized controlled trial, systematic reviews, technologies that collect large amounts of longitudinal data, etc. The strength of evidence needed is gauged by the user or purchaser such as payers (highly constrained) or the consumer (less constrained).[21]

**The need for information dissemination that bridges the gap between research and practice**

Currently there is a lack of awareness and knowledge about technologies among professionals, users and caregivers. Frequent dialogues between these three groups need to be established to discuss study designs and identify potential collaborations. They should be encouraged to identify factors and outcome markers for a success/effectiveness measurement of a technology. There is also a need for trusted sources of information such as a managed database of available devices (e.g., LeadingAge CAST) and facilitated access to technologies and the training/support to use them efficiently along with partnering between different technologies and understanding diverse attitudes towards devices and self-efficacy.

**The need for effective trans-disciplinary collaboration**

We need to develop opportunities for collaboration among experts in relevant disciplines (e.g., computer scientists, health researchers, clinicians, and engineers) that can integrate technology design, health literacy, and form factor expertise to develop and design technologies to meet the needs of an aging population, including the needs of specific subgroups, such as people with a particular chronic disease. We need to establish frequent dialogues between innovators, research funding agencies, and government regulatory agencies to discuss study designs and identify potential collaborations.

---

[21] Currently there is not a large evidence base supporting many promising technologies that could be widely used in practice. Classic telemedicine and telecare probably has the best or largest evidence base, although not all evidence has been positive (e.g., the Whole Systems Demonstration Project [McClean, 2013]). CMS does not reimburse for telemedicine based on current evidence. On the other hand, the VA has begun to widely incorporate telemedicine into its chronic care models for tertiary prevention.

**The need for far-reaching test beds**

We need to develop scalable 'test beds' in the community to efficiently, economically, and systematically explore the use of these technologies and involve the community in the research process and spur discovery science. These test beds need to accommodate different trajectories of care and heterogeneous sets of technologies. These test beds need novel funding mechanisms as they involve multiple aspects of research and development. Public and private support will be key.

**The need for patient access to actionable technologies**

Currently a tremendous barrier to the successful adoption of home based and mobile technologies is reimbursement through Medicare and Medicaid. Despite the limitations imposed by Medicare, the known benefits of remote patient monitoring and telehealth services include improved care, reduced hospitalizations, avoidance of complications, and improved satisfaction, particularly for the chronically ill.[22] There are also significant potential for cost savings; for example, remote monitoring is expected to result in savings of $36 billion globally by 2018, with North America accounting for 75% of those savings.[23] Although the newly released goal of HHS to move 50% of Medicare spending from FFS to value based care by 2018 will drive adoption of these technologies, a more direct approach in FFS for chronic disease patients is needed.

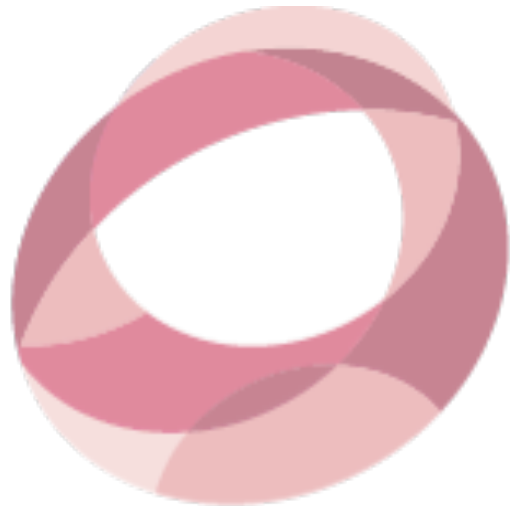

---

[22] *See, e.g.*, U.S. Agency for Healthcare Research and Quality (AHRQ) Service Delivery Innovation Profile, *Care Coordinators Remotely Monitor Chronically Ill Veterans via Messaging Device, Leading to Lower Inpatient Utilization and Costs* (last updated Feb. 6, 2013), *available at* http://www.innovations.ahrq.gov/content.aspx?id=3006.

[23] *See* Juniper Research, *Mobile Health & Fitness: Monitoring, App-enabled Devices & Cost Savings 2013-2018* (rel. Jul. 17, 2013), available at http://www.juniperresearch.com/reports/mobile_health_fitness.

**The following individuals participated in the workshop:**

| | | |
|---|---|---|
| Alicia | Anderson | U.S. Dept of Housing & Urban Development |
| Neeraj | Arora | National Cancer Institute |
| Stephen | Bauer | ED |
| Timothy | Bickmore | Northeastern University |
| Alice | Borrelli | Intel |
| Melinda | Buntin | Vanderbilt University |
| Margaret | Campbell | NIDRR |
| Neil | Charness | Florida State University |
| Octav | Chipara | University of Iowa |
| Elizabeth | Cocke | U.S. Dept of Housing & Urban Development |
| Lawton | Cooper | NHLBI |
| Theresa | Cruz | NICHD |
| Sara | Czaja | University of Miami |
| Susan | Czajkowski | National Heart, Lung, & Blood Institute |
| Anind | Dey | HCII, Carnegie Mellon University |
| Eric | Dishman | Intel |
| Sarah | Domnitz | Institute of Medicine |
| Ann | Drobnis | Computing Community Consortium |
| Thomas | Edes | Dept of Veterans Affairs |
| Jon | Eisenberg | NAS |
| Jerome | Fleg | NHLBI |
| Kenneth | Gabriel | Prognosys LLC |
| Gwendolyn | Graddy-Dansby | Center for Senior Independence |
| Dave | Gustafson | University of Wisconsin-Madison |
| Greg | Hager | Johns Hopkins University |
| Vicki | Hanson | Rochester Institute of Technology |
| Bill | Hanson | University of Pennsylvania Health System |
| Peter | Harsha | Computing Research Association |
| Robert | Hornyak | Administration for Community Living |
| Erin | Iturriaga | National Heart, Lung, and Blood Institute |
| Zack | Ives | University of Pennsylvania |
| Robert | Jarrin | Qualcomm Incorporated |
| Holly | Jimison | Northeastern University |
| Brian | Jones | Georgia Institute of Technology |
| Lyndon | Joseph | NIA |
| Emil | Jovanov | University of Alabama in Huntsville |
| Jeff | Kaye | ORCATECH Oregon Health & Science University |
| Steve | Kelly | Myomo |
| Jonathan | King | National Institute on Aging |
| Allison | Kumar | FDA/CDRH |
| Tony | Lee | Philips |
| Insup | Lee | University of Pennsylvania |
| Catherine | Levy | NHLBI |
| Clayton | Lewis | University of Colorado |
| Shari | Ling | CMS |
| Leah | Lozier | U.S. Dept of Housing & Urban Development |
| Chenyang | Lu | Washington University |
| Keith | Marzullo | NSF |
| Susan | Miller | CMS |
| Sandra | Mitchell | National Cancer Institute |
| Andrew | Mitz | NIH |
| Elizabeth | Mynatt | Georgia Institute of Technology |
| Wendy | Nilsen | National Institutes of Health |
| Misha | Pavel | Northeastern University |
| Andrew | Pope | Institute of Medicine |
| Louis | Quatrano | NCMRR/NICHD/NIH |
| Matthew | Quinn | FCC |
| Marilyn | Rantz | University of Missouri |
| Jamie | Roberts | NIH/NINDS |
| Mary | Rodgers | NIBIB |
| Wendy | Rogers | Georgia Institute of Technology |
| Jon | Sanford | Georgia Institute of Technology |
| Maureen | Schmitter-Edgecombe | Washington State University |
| Richard | Schulz | University of Pittsburgh |
| Weisong | Shi | NSF |
| Daniel | Siewiorek | Carnegie Mellon University |
| Nina | Silverberg | NIA |
| Margorie | Skubic | University of Missouri |
| Oleg | Sokolsky | University of Pennsylvania |
| Bob | Sproull | University of Pennsylvania |
| Mani | Srivastava | UCLA |
| John | Stankovic | University of Virginia |
| Carol | Star | U.S. Dept of Housing & Urban Development |
| Shar | Steed | Computing Research Association |
| Erika | Tarver | Foundation for the National Institutes of Health |
| Ranson | Towsley | Presbyterian SeniorCare |
| Helen | Vasaly | Computing Community Consortium |
| Howard | Wactlar | Carnegie Mellon University |
| Mary | Weick-Brady | US Food and Drug Administration |
| Michael | Weinrich | NICHD |
| Victoria | Zagaria | Intel Federal Healthcare |

This material is based upon work supported by the National Science Foundation under Grant No. (1136993).

Any opinions, findings, and conclusions or recommendations expressed in this material are those of the author(s) and do not necessarily reflect the views of the National Science Foundation, the National Heart, Lung, and Blood Institute or the National Institutes of Health.